\documentclass[letterpaper]{article} 
\usepackage{aaai25}  
\usepackage{times}  
\usepackage{helvet}  
\usepackage{courier}  
\usepackage[hyphens]{url}  
\usepackage{graphicx} 
\usepackage{natbib}  
\usepackage{caption} 
\usepackage{algorithm}
\usepackage{algorithmic}

\usepackage{booktabs}

\usepackage{array}

\usepackage{xcolor}

\usepackage{newfloat}
\usepackage{listings}
\DeclareCaptionStyle{ruled}{labelfont=normalfont,labelsep=colon,strut=off} 
\floatstyle{ruled}
\newfloat{listing}{tb}{lst}{}
\floatname{listing}{Listing}
\nocopyright

\title{The Effects of Moral Framing on Online Fundraising Outcomes: \\Evidence from GoFundMe Campaigns}

\author {
    Ji Eun Kim\textsuperscript{\rm 1},
    Libby Hemphill\textsuperscript{\rm 1}
}
\affiliations {
    \textsuperscript{\rm 1}School of Information, University of Michigan\\
    jieunkim@umich.edu, 
    libbyh@umich.edu
}

\usepackage{bibentry}

\begin{document}

\maketitle

\begin{abstract}
This study examines the impact of moral framing on fundraising outcomes, including both monetary and social support, by analyzing a dataset of 14,088 campaigns posted on GoFundMe. We focused on three moral frames: care, fairness, and (ingroup) loyalty, and measured their presence in campaign appeals. Our results show that campaigns in the Emergency category are most influenced by moral framing. Generally, negatively framing appeals by emphasizing harm and unfairness effectively attracts more donations and comments from supporters. However, this approach can have a downside, as it may lead to a decrease in the average donation amount per donor. Additionally, we found that loyalty framing was positively associated with receiving more donations and messages across all fundraising categories. This research extends existing literature on framing and communication strategies related to fundraising and their impact. We also propose practical implications for designing features of online fundraising platforms to better support both fundraisers and supporters.
\end{abstract}

%

\section{Introduction}

Online fundraising platforms have become an important resource for help seekers, allowing them to effectively communicate with their supporters and raise funds for various needs. Although these platforms host numerous members and campaigns, attention and support are not evenly distributed \cite{chakraborty2019impact}. Furthermore, the majority of campaigns fail to reach their goals. For example, only about 22\% of the campaigns in our dataset reached their goals, highlighting the challenge many fundraisers face in capturing sufficient attention and support on the platform. Therefore, it is important to identify the characteristics of successful fundraising campaigns to better assist help seekers.

\begin{figure}[t]
\centering
\includegraphics[width=0.9\columnwidth]{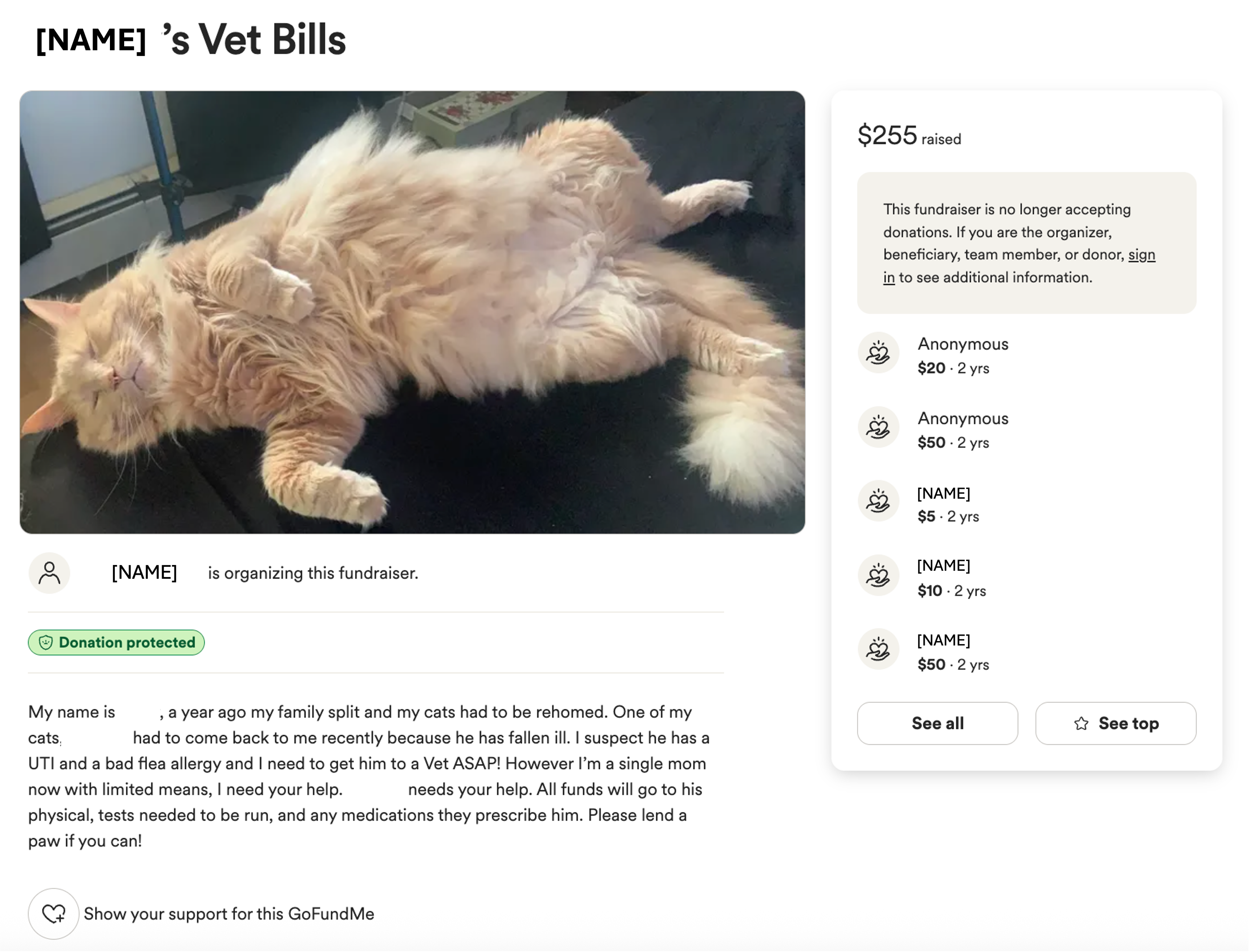}
\caption{Example GoFundMe campaign. Fundraisers can upload their stories with photos to provide background information and explain why they want to raise funds. Donors can donate money and leave support messages. The fundraiser and donor names have been removed from the screenshot.}
\label{fig1}
\end{figure}

This paper empirically examines how the moral framing of fundraising campaign appeals affects both monetary and social support outcomes. To achieve this goal, we model the moral framing of campaign appeals and measure its impact on fundraising outcomes by delving into the case of GoFundMe\footnote{https://www.gofundme.com/}. GoFundMe is one of the largest online fundraising platforms, providing a space where individuals facing various challenges seek financial and social support (see Figure \ref{fig1}). In 2024, GoFundMe received more than 42 million donations to individuals and over 23 million donations to non-profit organizations \cite{gofundme_1}. Donations occurred at an average rate of two per second on the platform, with an average amount of \$77. Fundraisers set a fundraising goal and post a story explaining the purpose and need of the campaign, along with photos. Donors can contribute financially to the campaign and have the option to donate anonymously. They can also leave supportive messages for the fundraiser. We analyzed a dataset of 14,088 GoFundMe campaigns to examine the impact of moral framing on the number of donations, the average donation amount per donor, and the number of support messages.

Our multiple linear regression analyses indicate that the impact of moral framing is strongest for campaigns in the Emergency category. Typically, negatively framed appeals that emphasize harm and unfairness attracted a larger number of donations and comments. However, for campaigns in the Memorial category, these appeals led to smaller average donation amounts per donor. Emphasizing ingroup loyalty was consistently effective in attracting more donations and comments across all fundraising categories.

This study offers both theoretical and practical contributions. Theoretically, our findings advance the current literature on donation behaviors and dynamics by examining how different moral frames impact campaign outcomes—both monetary and social support—across various fundraising categories. Practically, these insights can inform the design of support tools that assist fundraisers in crafting effective appeals tailored to specific situations and goals. 

\section{Related Work}

\subsection{Determinants of Successful Online Fundraising}

According to \citet{hoover2018moral}, promotive factors affecting the success of charitable donations can be categorized into three types: individual differences, situational characteristics, and solicitation framing. While previous studies have demonstrated that the first two types significantly influence the success of online fundraising, this research focuses specifically on examining the impact of solicitation framing, particularly in the context of text-based campaign appeals.

First, individual differences relate to donor characteristics and the shared group identities between givers and recipients that influence donation decisions. For example, on the GoFundMe platform, both men and women are more likely to donate when the majority of supporters are of the opposite sex, and female supporters often express greater empathy in their support messages \cite{sisco2019examining}. Recipient characteristics also play a role because donors tend to give more to those with the same last names \cite{sisco2019examining} and prefer supporting ingroup members over outgroup ones \cite{nilsson2016congruency}. Additionally, female-led campaigns generally achieve their fundraising goals more quickly \cite{esperancca2025determinants}.

The second category is situational characteristics, which include factors like social pressure and reputation-seeking. Research has shown that people engage in helping behaviors to achieve a ``warm glow'' from helping others and to avoid the guilt associated with not helping \cite{cunningham1980wanting}. These motivations are influenced differently by situational factors, such as the level of personal responsibility. For instance, individuals are more affected by a sense of guilt in high-responsibility situations, as they feel guiltier for not helping when under pressure. Conversely, in low-responsibility situations, they tend to experience a warm glow, possibly because helping feels like an additional, voluntary action \cite{erlandsson2016anticipated}. Other studies suggest that people seek to enhance their reputation through helping behaviors. For example, they may increase their donation amounts when the average donation amount is visible \cite{sisco2019examining} and imitate high-status individuals who donate \cite{kumru2010effect}.

Finally, solicitation framing involves crafting appeals that encourage more donations and social support. Understanding these strategies requires examining both image- and text-based information in campaign appeals. Some researchers have focused on image-based features, analyzing photos included in campaign appeals to measure their effects on fundraising performance. For example, \citet{rhue2018emotional} found that campaigns featuring photos with happy facial expressions tend to collect more donations, whereas neutral facial expressions result in fewer donations. Another study assessed the gender, age, emotional state, and attractiveness of help seekers in photos attached to GoFundMe campaigns. It also evaluated the aesthetic and technical quality of these images to incorporate these factors into machine learning models predicting fundraising outcomes \cite{zhang2021contributes}. On the other hand, other researchers have examined text-based information that contributes to successful campaigns, which is the focus of this research. For instance, longer campaign appeals have been found to help achieve fundraising goals more quickly \cite{esperancca2025determinants}. Additionally, research indicates that using more netspeak in appeals can positively influence fundraising outcomes, while excessive use of second-person pronouns may have adverse effects \cite{jang2022effect}. The next subsection delves into the existing literature on the language and framing used to create effective fundraising appeals.

\subsection{The Framing of Fundraising Appeals}

Framing is an effective tool for persuasion, and several framing strategies have been adopted to encourage charitable donations \cite{wang-etal-2019-persuasion}. It is especially crucial in fundraising because seeking financial assistance can be stigmatizing and often leads to criticism \cite{parsell2022charity}. This stigma makes it challenging for vulnerable and stigmatized groups to ask for help. Consequently, previous research has explored how fundraisers frame their stories to justify their requests without degrading themselves or creating negative impressions. For instance, \citet{radu2018qualitative} found that victims of intimate partner violence frame their stories carefully by portraying themselves positively to counteract the social stigma associated with seeking financial help. Their strategies include framing their issues as part of the broader issue of domestic violence to evoke a sense of social justice and including photos of their innocent children to justify their requests for support.

Existing work on fundraising appeals primarily focuses on emotions as crucial components for framing appeals. Research has shown that both positive and negative emotions can be beneficial. Positively framed appeals often evoke a sense of warm glow, while negatively framed ones can induce feelings of guilt \cite{erlandsson2018attitudes}. However, the findings from previous studies are mixed and nuanced, as the effects of emotions can vary based on the fundraising context, goals, and outcomes. Some research highlights the effectiveness of positive emotional appeals; for instance, positive sentiments in campaigns aimed at promoting social bonding have been shown to generate more donations \cite{paxton2020does}. Another study suggests that tweets emphasizing positive moral values, such as care and loyalty, encourage donations \cite{hoover2018moral}. Conversely, another line of research indicates that negative emotions might be more effective in promoting helping behavior. For example, some researchers emphasize the power of anger, as people are inclined to donate more to restorative charities aiming for equity \cite{blader2002justice, van2017impact}. Another study suggests that joyful language may lead to smaller donation amounts \cite{rhue2018emotional}. \citet{jang2022effect} found that although negativity might attract more donors and social media attention, it could result in smaller donation amounts per donor.

\subsection{Research Gaps and Questions}

We aim to address the following research gaps in the current literature. First, previous studies have produced mixed findings because the impacts of text-based features can vary depending on different fundraising purposes and needs. Therefore, we include various fundraising categories in our analysis to investigate how moral framing affects each category differently. Second, while previous research has examined how overall emotional framing influences fundraising outcomes, it often relied on a binary opposition of positive and negative sentiments, which may be too simplistic. We aim to extend this line of research by using natural language processing methods to extract more nuanced linguistic features from campaign appeals. Although moral frames we use are also based on binary oppositions (e.g., harm versus care), we use an ensemble of three binary oppositions to represent multiple frames in a parsimonious way. We investigate how the moral framing of campaign appeals affects the following outcomes: (1) the number of donations, (2) the average donation amount per donor, and (3) the number of comments from supporters.

Our first research question explores how differently framed campaign appeals influence varying donation outcomes. Based on the literature, we anticipate that negative framing will attract a greater number of donations but result in smaller amounts per donor. Our second research question examines whether the number and content of comments a campaign receives vary according to the moral framing of the appeal.



\begin{itemize}
  \item \textbf{RQ1}: Do differently framed campaign appeals result in varying donation outcomes?
  \item \textbf{RQ2}: Do differently framed campaign appeals influence both the quantity and quality of social support?
\end{itemize}


\section{Data and Methods}

\subsection{Data}

We utilized a publicly available dataset\footnote{https://zenodo.org/records/8287320} consisting of 14,859 unique GoFundMe campaigns uploaded in 2022 \cite{xu2023mdcc}. While our dataset is publicly available, we aim to prevent potential harm by adopting several strategies, such as data aggregation and paraphrasing. Our results are presented in aggregated form, and any representative quotes used for illustrative purposes are paraphrased to further safeguard user privacy. The dataset includes 13 unique countries supported by the platform, but most campaigns were posted by U.S.-based fundraisers. We excluded campaigns posted by non-U.S.-based fundraisers and those with non-English appeals using Google's language detection library, \textit{langdetect}. This process resulted in a total of 14,088 campaigns.

The fundraising campaigns in our final dataset were initially categorized into five main categories: Memorial (30.6\%), Medical (23.7\%), Animals (19.5\%), Emergency (16.9\%), and Financial Emergency (9.3\%). We then merged Emergency and Financial Emergency into a single category (Emergency), resulting in four categories. The total number of donation transactions is 1,142,406, and the total number of comments is 52,550. About 22\% of the campaigns achieved their fundraising goals.

\subsection{Measuring Moral Values Using Word Embeddings}

We adopt the following three moral frames proposed by Moral Foundations Theory (MFT): care, fairness, and loyalty. The original MFT identifies five moral foundations (care-harm, fairness-cheating, loyalty-betrayal, authority-subversion, and sanctity-degradation) to explain variations in moral judgment and development \cite{graham2013moral, haidt2007morality}. These moral foundations have been widely used to analyze how moral judgments influence morally relevant outcomes and have been applied to investigate their impacts on charitable giving behaviors \cite{nilsson2016congruency, hoover2018moral}.

The moral foundation of \textit{care} involves protecting people from harm, while \textit{fairness} pertains to the human need for equal treatment. The \textit{loyalty} foundation relates to group formation and belonging. The \textit{authority} foundation concerns respecting authority and tradition, and \textit{sanctity} involves avoiding contamination and living in a spiritually clean manner. Among these five moral foundations, we chose care, fairness, and loyalty for this analysis for two reasons. First, previous research suggests that these three frames can influence fundraising outcomes \cite{blader2002justice, hoover2018moral, linos2021fundraising, nilsson2016congruency, van2017impact}. Additionally, authority is conceptually linked to relationships with higher authorities, and sanctity is more relevant to religious contexts, indicating that these two frames may not be well-suited for online fundraising.

\begin{table*}[t]
\centering
\setlength{\tabcolsep}{8pt}
\renewcommand{\arraystretch}{1.1}
\begin{tabular}{llll}
    \toprule
    Moral domain & Bias & Number of terms & Example seed terms\\
    \midrule
    Care & Vice &74 & abuse, attack, damage, destroy, harm \\
     &  Virtue &57 & benefit, care, defend, empathic, peace\\
    \midrule
    Fairness & Vice &39 & discriminate, dishonest, injustice, prejudice, unfair \\
    & Virtue & 38 & equity, fair, honest, justice, reciprocal \\
   \midrule
    Loyalty & Vice & 29 & deceive, enemy, foreign, immigrant, imposter \\
    & Virtue & 59 & ally, community, family, fellow, group \\
    \bottomrule
\end{tabular}
\caption{Moral frames adopted for this study and example seed words representing each semantic construct.}
\label{table1}
\end{table*}

We used the code from \citet{mokhberian2020moral} to computationally construct moral frames. \citet{mokhberian2020moral} employed the FrameAxis approach \cite{kwak2021frameaxis} to construct moral axes and compute bias scores for news headlines and articles on each axis. This method involves measuring the cosine similarity between the words in a text and the axis to determine bias. The FrameAxis approach has been used to measure moral expressions in social media posts related to social movements \cite{priniski2021mapping} and partisanship in political news \cite{mokhberian2020moral}. Each axis consists of two opposing poles represented by antonym pairs, and we used words from the moral foundations dictionary developed by \citet{graham2009liberals} to define each end of the axis. For example, the harm pole of the harm-care axis consists of harm-related words, such as “abuse” and “damage,” while the care pole is defined using care-related words like “compassion” and “safe.” The number of terms and examples included in each moral frame can be found in Table \ref{table1}. We calculate the harm-care axis (i.e., care frame) by subtracting the average of the word vectors of care-related terms from the average of the word vectors of harm-related terms. We obtained the word vectors from the GloVe\footnote{https://nlp.stanford.edu/projects/glove/} word embedding model trained on a large Twitter corpus. 

After constructing the moral axes, which represent different moral frames, we characterized each campaign appeal by measuring its bias toward these axes. For example, the bias score of a campaign appeal on the care dimension is calculated by taking the weighted average of the cosine similarities between its words and the axis. If a campaign appeal contains many words with high cosine similarity to the care pole, it is considered more relevant to the care pole than to the harm pole. A positive bias score indicates that the campaign appeal is more aligned with the care pole, while a negative score suggests it contains more harm-related words (see Table \ref{table2} for examples of campaign appeals with varying moral foundation scores).

\begin{table*}[t]
\centering
\setlength{\tabcolsep}{8pt}
\renewcommand{\arraystretch}{1.1}
\begin{tabular}{ll >{\raggedright\arraybackslash}p{11cm}}
    \toprule
    Moral domain & Bias & Campaign appeal\\
    \midrule
    Care & Vice & Last night my bunny was attacked by a neighbor's cat, which somehow managed to open the cage and bite him. [...] \\
     & Virtue & We are collecting funds to cover essential legal representation and travel expenses to secure my daughter's physical, mental, and emotional safety. [...]\\
    \midrule
    Fairness & Vice & I'm raising funds for [NAME], who tragically lost his life due to a crime over a phone at a train station. [...] \\
     & Virtue & This family has faced numerous challenges without giving up, but they are now under pressure to save their home. [...] No one deserves to lose their home. \\
   \midrule
    Loyalty & Vice & Offering shelter and medication to stray animals. \\
     & Virtue & In this difficult time, it's important for us to come together as a community to support [NAME]’s family. [...] \\
    \bottomrule
\end{tabular}
\caption{Examples of campaign appeals with varying moral foundation scores. Note that excerpts from campaign appeals presented above have been paraphrased due to privacy concerns.}
\label{table2}
\end{table*}

\subsection{Variables}

The outcome variables of this study include the number of donations, the average donation amount per donor, and the number of comments. We excluded the total amount of money raised from this analysis because it is highly correlated with the number of donations (Spearman’s $\rho$ = 0.91, \textit{p} $<$ .001). We conducted multiple linear regression analyses to examine the main and interaction effects of fundraising categories and moral foundation scores on each continuous outcome variable. The moral foundation scores for three moral frames were calculated using the methods described in the previous subsection.

We added several control variables to our models that may impact fundraising outcomes and social support. First, we used VADER, a sentiment analysis tool \cite{hutto2014vader}, to label the sentiment scores of campaign appeals in our dataset. Each appeal was classified into one of three groups—positive, neutral, or negative—based on its compound score as determined by VADER. Most of the campaign appeals in our dataset were positive; the percentages of positive, neutral, and negative appeals were 87.8\%, 0.7\%, and 11.4\%, respectively. We also measured the length of campaign appeals by splitting each text by whitespace and counting the number of tokens. Additionally, we included the number of photos in an appeal and the fundraising goal amount as control variables. The descriptive statistics for the continuous variables in our analysis are presented in Table \ref{table3}.

\begin{table*}[t]
\centering
\setlength{\tabcolsep}{8pt}
\renewcommand{\arraystretch}{1.1}
\begin{tabular}{lcccc}
    \toprule
     & Animals & Emergency & Medical & Memorial\\
    \midrule
    Number of donations & 31.51 (39.34) & 57.27 (88.31) & 123.95 (135.11) & 99.91 (127.14)\\
    Average donation amount per donor & 53.89 (34.62) & 86.89 (54.71) & 107.32 (55.77) & 96.12 (50.50)\\
    Number of comments & 1.40 (2.93) & 2.27 (4.79) & 5.77 (9.09) & 4.88 (7.46)\\
    \midrule
    Care score & -0.06 (0.02) & -0.05 (0.02) & -0.05 (0.02) & -0.06 (0.02)\\
    Fairness score & 0.23 (0.02) & 0.24 (0.03) & 0.22 (0.03) & 0.23 (0.03)\\
    Loyalty score & 0.10 (0.02) & 0.11 (0.02) & 0.10 (0.02) & 0.13 (0.02)\\
     \midrule
       Campaign appeal length & 225.73 (182.26) & 218.38 (205.69) & 285.76 (233.43) & 159.16 (116.69)\\
       Number of photos & 0.74 (2.27) & 0.36 (1.48) & 0.39 (1.49) & 0.23 (1.41)\\
    Fundraising goal & 4.7k (9.1k) & 12.0k (29.1k) & 327.9k (17.3M) & 14.2k (19.0k)\\
    \bottomrule
\end{tabular}
\caption{Descriptive statistics for our dataset. The numbers in the cells are mean values, and those in parentheses are standard deviations.}
\label{table3}
\end{table*}

\begin{table*}
\centering
\setlength{\tabcolsep}{8pt}
\renewcommand{\arraystretch}{1.1}
\begin{tabular}{p{3.5cm} p{3.6cm} p{3.6cm} p{3.6cm}}
    \toprule
     &\textbf{Model 1}&\textbf{Model 2}&\textbf{Model 3}\\
    \midrule
    Outcome variable & Number of donations & Average donation amount & Number of comments\\
    \midrule
    Emergency & 0.951*** & 0.267* & 0.330\\
  &(0.162) & (0.104) & (0.190)\\
    Medical & 0.733*** & 0.361***& 0.168\\
      & (0.164) & (0.105) & (0.192)\\
    Memorial & 1.211*** & 0.291** &0.365*\\
  & (0.156) &(0.100) &(0.183)\\
    Care &2.098**&2.122***&0.027\\
      &(0.799) &(0.512)&(0.938)\\
    Fairness & 0.446 &-0.962& -0.201\\
      & (0.812) &(0.520)& (0.954)\\
      Loyalty&-1.156 &1.314*&0.102\\
      &(0.982) &(0.629)&(1.153)\\
      \midrule
    Emergency $\times$ Care & -3.412*** &-0.228
& -2.563*\\
      & (1.015) & (0.650)& (1.192)\\
    Medical $\times$ Care & -0.377 &0.222 &0.948\\
      & (1.085) & (0.695)& (1.274) \\
    Memorial $\times$ Care & -1.011 & 1.018 &0.714\\
      & (1.028) & (0.658) & (1.208) \\
      Emergency $\times$ Fairness & -8.309***  &-0.472& -3.789**\\
      & (0.984) &(0.630)& (1.156)\\
    Medical $\times$ Fairness & -1.587& 0.934 &1.140\\
      & (1.022) &(0.654) & (1.200) \\
    Memorial $\times$ Fairness & -3.706*** & 1.175*&-0.614\\
      & (0.924)&(0.592) & (1.086) \\
      Emergency $\times$ Loyalty & 9.252*** &1.666*& 4.180**\\
      & (1.229)&(0.787)& (1.444)\\
    Medical $\times$ Loyalty & 4.319*** &  -0.980 &1.413\\
      & (1.262)&(0.808)& (1.482) \\
    Memorial $\times$ Loyalty & 3.005* & -0.600 &2.810*\\
      & (1.203)& (0.770)& (1.413) \\
      \midrule
    Positive sentiment & -0.005& 0.029* & 0.026\\
      &(0.019) & (0.012)&  (0.022)\\
   Neutral sentiment & -0.092& 0.061&-0.104\\
      &(0.070) &(0.045)& (0.082) \\
    Campaign appeal length & 0.051***& 0.027***&0.078***\\
      & (0.009)&(0.006) &  (0.010) \\
    Number of photos & 0.033** & 0.006&0.017 \\
      & (0.013)& (0.008)& (0.015)  \\
      Fundraising goal & 0.281*** &0.123***& 0.206***\\
      &(0.005) & (0.003)& (0.006) \\
    \midrule
    Intercept & 0.825***&2.950*** & -1.444***\\
      & (0.143)&(0.091) &  (0.168) \\
     \midrule
    Number of observations & 14,088 & 14,088 & 14,088 \\
    Adjusted $R^2$ & 0.45 & 0.31 & 0.23\\
    \bottomrule
\end{tabular}
\caption{Multiple linear regression results with the outcome variable for Model 1 being the number of donations, for Model 2 being the average donation amount per donor, and for Model 3 being the number of comments. The numbers in the cells represent coefficients, and those in parentheses are standard errors. $^{*}p<0.05$; $^{**}p<0.01$; $^{***}p<0.001$.}
\label{table4}
\end{table*}

\section{Results}

We conducted multiple regression analyses to address our research questions, ensuring that our dataset met all the necessary assumptions for regression analysis. We included two categorical variables—fundraising category and campaign appeal sentiment—using ``Animals'' and ``Negative'' as the reference categories, respectively, across all models. A log transformation was applied to the following variables: number of donations, average donation amount per donor, number of comments, campaign appeal length, number of photos, and fundraising goal amount. Table \ref{table4} presents the outputs of three multiple regression models, with the outcome variable for Model 1 being the number of donations, for Model 2 the average donation amount per donor, and for Model 3 the number of comments.

For our control variables, both the campaign appeal length and the fundraising goal amount were positively associated with all outcome variables. Positive sentiment, however, was positively associated only with the average donation amount per donor (\textit{b} = 0.029, \textit{p} $<$ .05) and did not have a significant impact on the number of donations or comments. Additionally, the number of donations increased with the number of photos included in an appeal (\textit{b} = 0.033, \textit{p} $<$ .01).

\subsection{The Impact of Moral Framing on the Number of Donations}

Model 1 in Table \ref{table4} presents the effects of three moral frames on the number of donations, along with their interactions with different fundraising categories. The main effect of care framing, which is the impact of care framing on the Animals category, is significant and positively associated with the number of donations (\textit{b} = 2.098, \textit{p} $<$ .01). However, in the Emergency category, campaigns with higher care scores received fewer donations (\textit{b} = -3.412, \textit{p} $<$ .001), suggesting that campaigns highlighting harm collected more donations. Similarly, fairness framing is negatively associated with the number of donations in the Emergency (\textit{b} = -8.309, \textit{p} $<$ .001) and Memorial (\textit{b} = -3.706, \textit{p} $<$ .001) categories, implying that campaigns emphasizing unjust or unfair situations generate more donations. In contrast, campaigns that emphasize ingroup loyalty received more donations. Higher loyalty scores increased the number of donations in the Emergency (\textit{b} = 9.252, \textit{p} $<$ .001), Medical (\textit{b} = 4.319, \textit{p} $<$ .001) and Memorial (\textit{b} = 3.005, \textit{p} $<$ .05) categories.   

\subsection{The Impact of Moral Framing on the Average Donation Amount per Donor}

Model 2 in Table \ref{table4} assesses the effects of three moral frames on the average donation amount per donor, along with their interaction with different fundraising categories. In the Animals category, a one-unit increase in the care score is associated with an average increase of \$2.122 in the donation amount per donor, after controlling for other variables (\textit{b} = 2.122, \textit{p} $<$ .001). However, care framing did not show significant interaction effects in other categories. In the Memorial category, campaigns with higher fairness scores resulted in higher average donation amounts per donor (\textit{b} = 1.175, \textit{p} $<$ .05). Additionally, the effect of loyalty framing is significant and positively associated with the average donation amount per donor in the Animals category (\textit{b} = 1.314, \textit{p} $<$ .05), with an even stronger positive impact in the Emergency category (\textit{b} = 1.666, \textit{p} $<$ .05).

\begin{figure*}[t]
\centering
\includegraphics[width=0.9\textwidth]{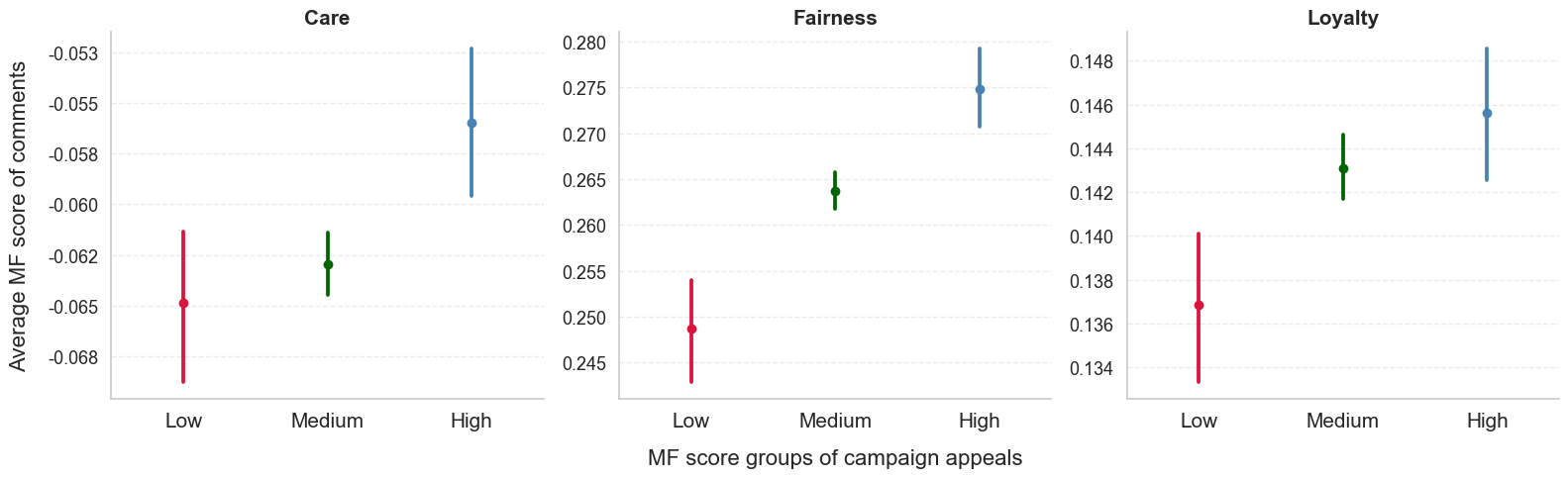}
\caption{Average moral foundation scores of comments for different score groups of campaign appeals across three moral frames. This figure specifically uses campaign appeals and their comments from the Emergency category.}
\label{fig2}
\end{figure*}

\begin{figure*}[t]
\centering
\includegraphics[width=0.9\textwidth]{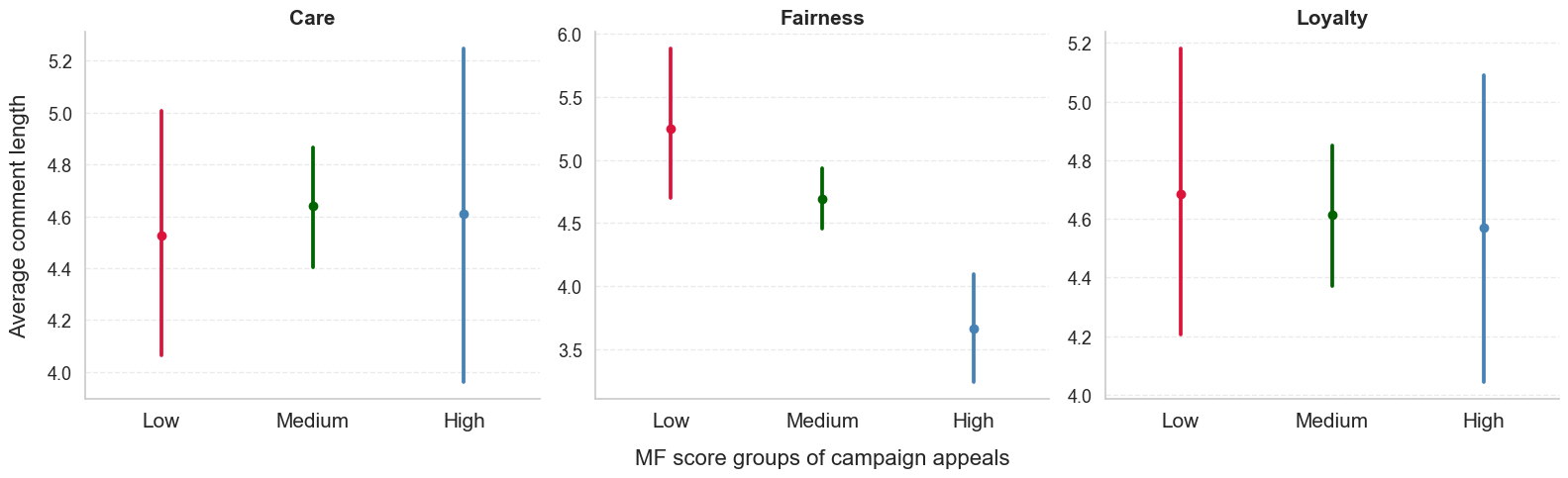}
\caption{Average comment length for different score groups of campaign appeals across three moral frames. We computed the length of each comment by splitting the text by whitespace and counting the resulting number of tokens. This figure specifically uses campaign appeals and their comments from the Emergency category.}
\label{fig3}
\end{figure*}

\subsection{The Impact of Moral Framing on Social Support}

Model 3 in Table \ref{table4} reports the effects of three moral frames on the number of comments from supporters, along with their interactions with different fundraising categories. Our results indicate that campaigns in the Emergency category attracted more comments when their appeals were negatively framed in terms of care (\textit{b} = -2.563, \textit{p} $<$ .05) and fairness (\textit{b} = -3.789, \textit{p} $<$ .01). The positive impact of loyalty framing was statistically significant in both the Emergency (\textit{b} = 4.180, \textit{p} $<$ .01) and Memorial (\textit{b} = 2.810, \textit{p} $<$ .05) categories, resulting in an increase in the number of comments. 

Since negatively framed campaign appeals in terms of care and fairness attract more support messages, we further investigated whether supporters post more positive comments in response to these appeals to compensate for or dilute negativity and encourage fundraisers. Given that the impact of moral framing was strongest in the Emergency category, we divided campaign appeals from this category into three groups based on their moral foundation score distribution for each moral frame: low (below the mean minus one standard deviation), medium (within one standard deviation of the mean), and high (above the mean plus one standard deviation). We also computed the moral foundation scores of all comments and calculated the average comment moral foundation score by aggregating these scores for each campaign appeal. Figure \ref{fig2} illustrates that, across all moral frames, comments tend to reflect the moral foundation scores of their corresponding campaign appeals. This means that appeals with lower moral foundation scores attract comments with lower scores, while those with higher scores tend to receive comments with higher scores. Thus, it appears that supporters do not use comments as a means to counterframe negatively framed appeals. However, Figure \ref{fig3} illustrates that negatively framed appeals highlighting injustice and unfairness receive significantly longer comments than positively framed ones. This suggests that the perceived justice of a situation may impact the quality of social support from supporters.

\section{Discussion}

\subsection{Main Findings}

Understanding how moral framing influences people to engage in prosocial behavior is essential for promoting online fundraising and social support for those seeking help. Our study aimed to examine how three moral frames—care, fairness, and (ingroup) loyalty—affect the donations and social support that GoFundMe fundraisers receive across different fundraising categories. Overall, our findings suggest that negatively framing appeals in terms of care and fairness, along with highlighting ingroup identities, could be effective for fundraising.

First, our results indicate that moral frames have varying effects on GoFundMe donation campaigns, depending on the fundraising categories and outcomes. In the Emergency category, care and fairness frames significantly influenced both monetary and social support outcomes. Specifically, negatively framed appeals attracted a greater number of donations and comments. One possible explanation for this is that supporters may have perceived these situations as more dangerous, urgent, or unjust, prompting a stronger reaction and resulting in rage donations \cite{van2017impact}. We also investigated whether supporters leave more positive support messages to counterframe or dilute the negativity present in campaign appeals. However, Figure \ref{fig2} demonstrates that the moral foundation scores of comments generally align with those of the campaign appeals. This suggests that supporters tend to help fundraisers by increasing their donation frequency rather than writing positive messages, even after encountering negatively framed appeals. Although supporters do not frame their messages positively, they tend to leave longer messages for appeals that are negatively framed in terms of fairness (see Figure \ref{fig3}).

\begin{figure}[t]
\centering
\includegraphics[width=1\columnwidth]{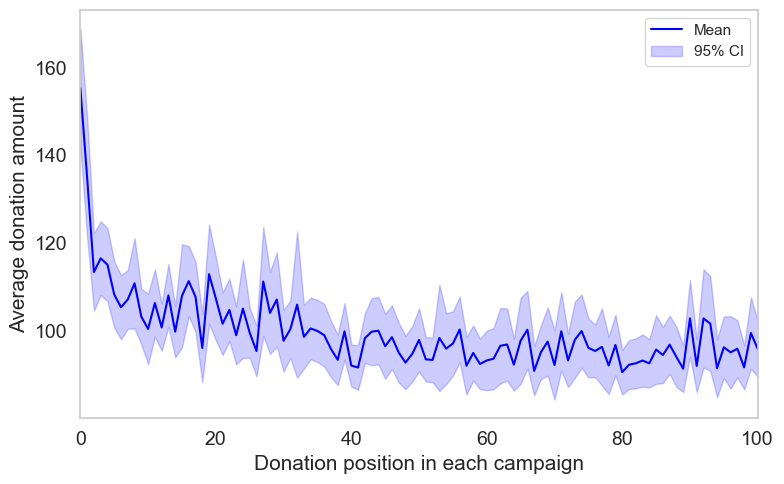}
\caption{Changes in average donation amount by donation position within each campaign. The line represents the mean donation amount at each sequential position, and the shaded area indicates the 95\% confidence interval. We included only campaigns that collected at least 100 donations for this plot.}
\label{fig4}
\end{figure}

Another interesting finding is that campaign appeals in the Memorial category earned a greater number of donations, but the average donation amount decreased when appeals were negatively framed in terms of fairness. In other words, negatively framed appeals might boost the total number of donations, but this can reduce the average donation amount per donor. This aligns with the findings of \citet{jang2022effect}, which suggest that negative emotional words in campaign appeals increase the number of donations while decreasing the contribution per donor. One possible explanation provided by \citet{jang2022effect} is that people tend to avoid negative feelings caused by not helping others. In such cases, individuals donate to alleviate these negative feelings but often contribute the minimum amount necessary to relieve the guilt. This behavior leads to a higher number of donations but smaller donation amounts per donor. Another possibility we propose is the diffusion of responsibility, as described by \citet{tsvetkova2015contagion}. In this scenario, individuals donate smaller amounts when they perceive their contributions as less critical, given the large number of other donors, as they believe fundraisers are already receiving sufficient help. Figure \ref{fig4} illustrates this trend by showing the average amount given at the first donation, second donation, third donation, and so on, aggregated across campaigns. The trend indicates that the average donation amount decreases over time, supporting the diffusion of responsibility theory.

Unlike care and fairness framing, loyalty framing generally proved beneficial for attracting both monetary and social support when positively framed. Several studies have shown that loyalty framing is more effective for fundraising since people are more inclined to help those they perceive as similar to themselves rather than outgroup members \cite{nilsson2016congruency, sisco2019examining}. However, the consistently positive impact of ingroup loyalty on social support warrants further investigation, as it contradicts research suggesting that expressing positivity reduces the amount of social support help seekers receive, since supporters may feel less compelled to provide such support \cite{rhue2018emotional, walsh2021can}.

\subsection{Implications}

The findings of this research have several theoretical implications. First, this study extends existing work on the impact of moral framing on donation outcomes. Beyond examining the impact of a single binary opposition between positive and negative sentiments, we measured the distinct effects of the care, fairness, and loyalty dimensions. This approach provides nuanced insights into the diverse ways moral language can motivate donations and social support. Additionally, our study suggests that the effects of moral framing may vary across different fundraising purposes and situations. Furthermore, the findings regarding supporters’ commenting behaviors raise new questions about how supporters engage with moral frames. It appears that instead of counterframing the negativity reflected in campaign appeals with positive messaging, users choose to increase their donation frequency. This behavior might indicate a compensation mechanism where financial support is preferred over changing the discourse.

The findings of this research have several practical implications, particularly for campaign strategies and the design of fundraising activities and platforms. First, this study provides insights into what constitutes an effective fundraising appeal. Fundraisers can tailor their message framing according to their specific purposes and goals. In emergency situations, leveraging negative framing related to care and fairness may attract more support by highlighting harm or injustice. Meanwhile, emphasizing ingroup loyalty can be effective across all types of campaigns. Furthermore, these findings suggest potential features for fundraising platforms to optimize framing effectiveness. Platforms could offer tools or recommendations to assist fundraisers in framing their campaigns, along with analytics to track the impact of different framing strategies on donations and engagement. Given that the moral foundation scores of comments align with those of appeals, platforms may try enhancing their social support systems. For example, they can explore prompting differently framed messages to guide supporters in writing helpful comments, thereby enhancing emotional support.

\subsection{Limitations and Future Work}

Here, we discuss some limitations of this work and highlight opportunities for future research. The first limitation concerns the generalizability of the research, as our dataset includes only campaign appeals and comments written in English and posted by U.S.-based fundraisers. Additionally, the dataset is limited by a lack of detailed information about fundraisers and supporters on the platform (e.g., gender, social status), factors that have been shown to affect fundraising outcomes. Securing such data could enable future research to perform propensity score matching, allowing for the identification of comparable campaigns to measure the causal impact of moral framing on fundraising outcomes.

Another limitation of this study is the difficulty in pinpointing the exact mechanisms behind our observations. Although we conducted several follow-up analyses to propose potential explanations for our findings, this research still raises questions about donation behaviors and dynamics. For instance, after observing that negatively framed appeals related to care and fairness increased the quantity of comments from supporters, we examined the quality of these comments. We initially hypothesized that negatively framed appeals would attract more positive comments from supporters seeking to counterbalance the negativity. However, our analysis did not support this hypothesis. This underscores the need for a deeper examination of the quality and content of social support provided by donors. Future research could benefit from carefully designed experiments or surveys to explore these dynamics in greater detail.

Lastly, we computationally modeled the language and framing of campaign appeals using existing resources, such as MFT, which represents moral frames in the form of binary oppositions, and a corresponding dictionary. While these frames have been widely used across various contexts and disciplines, future research could develop customized dictionaries and novel dimensions that are better suited to online fundraising contexts. For instance, \citet{best2023stigma} constructed four stigma dimensions—immorality, negative personality traits, disgust, and danger—to examine how diseases have been depicted negatively in major newspapers over time. In addition to moral dimensions, incorporating socioeconomic dimensions such as class, gender, and race, as adopted by \citet{kozlowski2019geometry}, could be valuable for future research. These dimensions have been shown to influence donation dynamics and outcomes, and they may offer deeper insights into how the positioning of campaign appeals along these dimensions affects fundraising success.

\section{Conclusion}

This research investigates how the moral framing of GoFundMe campaign appeals influences the monetary and social support that fundraisers receive. We measured the moral foundation scores of three moral frames (care, fairness, and loyalty) for each campaign appeal using word embeddings. Among various categories, the fundraising outcomes in the Emergency category were most sensitive to moral framing. Our findings suggest that negatively framing campaign appeals in terms of care and fairness can effectively attract a greater number of donations and comments. However, such negative framing may backfire in the Memorial category by reducing the average donation amount per donor. Additionally, we found that emphasizing ingroup loyalty is beneficial for securing more monetary and social support across all categories.

\bibliography{aaai25}

\end{document}